\documentclass[10pt,final,conference,a4paper,oneside,twocolumn]{IEEEtran}
\usepackage{amsmath}
\usepackage{graphicx}
\usepackage{amssymb}
\newcommand{\Mod}[1]{\ (\text{mod}\ #1)}

\begin{document}

\title{Cold-atom Inertial Sensor without Deadtime}
\author{\IEEEauthorblockN{B. Fang, I. Dutta, D. Savoie, B. Venon, C. L. Garrido Alzar, R. Geiger and A. Landragin}
\IEEEauthorblockA{LNE-SYRTE, Observatoire de Paris, PSL Research University, CNRS, Sorbonne Universit\'es\\
  UPMC Univ. Paris 06, 61 avenue de l’Observatoire, 75014 Paris, France\\
  Email: bess.fang@obspm.fr}}
\maketitle

\begin{abstract}
We report the operation of a cold-atom inertial sensor in a joint
interrogation scheme, where we simultaneously prepare a cold-atom
source and operate an atom interferometer in order to eliminate dead
times.  Noise aliasing and dead times are consequences of the
sequential operation which is intrinsic to cold-atom atom
interferometers.  Both phenomena have deleterious effects on the
performance of these sensors.  We show that our continuous operation
improves the short-term sensitivity of atom interferometers, by
demonstrating a record rotation sensitivity of
$100$~nrad.s$^{-1}/\sqrt{\rm Hz}$ in a cold-atom gyroscope of
$11$~cm$^2$ Sagnac area.  We also demonstrate a rotation stability of
$1$~nrad.s$^{-1}$ after $10^4$ s of integration, improving previous
results by an order of magnitude.  We expect that the continuous
operation will allow cold-atom inertial sensors with long
interrogation time to reach their full sensitivity, determined by the
quantum noise limit.
\end{abstract}

\section{Introduction}
Over the last twenty years, inertial sensors based on atom
interferometry have evolved significantly in terms of performance and
transportability.  Such progress ensures the relevance of atom
interferometer (AI) based inertial sensors in various field
applications, ranging from inertial
navigation~\cite{Canuel2006,Geiger2011,stockton_absolute_2011} to
geophysics and
geodesy~\cite{Gillot2014,Geiger2015,Freier2015,RosiMeasurement2015},
as well as in fundamental
physics~\cite{AltschulQuantum2015,Dimopoulos2008PRD,Chaibi2016}.
Although new techniques are currently explored to further improve the
sensitivity of these
sensors~\cite{Clade2009,Chiow2011,Debs2011,Leroux2010,Hosten2016}, the
issues associated with measurement dead time remain a strong
obstacle to their ultimate performance~\cite{Fang2016}.

Dead times in AIs correspond to the time needed to prepare and to
detect the atoms before and after the interferometric sequence.  They
result in loss of inertial information, leaving AIs unsuitable for
inertial measurement units (IMUs) in
navigation~\cite{jekeli_navigation_2005} or for recording fast varying
signals in seismology~\cite{Schreiber2013} for the time being.  Noise
aliasing coming from the sequential operation also degrades the AI
sensitivity in the presence of dead times, similar to the Dick effect
in cold atomic clocks~\cite{SantarelliFrequency1998}.  

In this paper, we report the first continuous operation of a cold-atom
inertial sensor.  This is demonstrated in a gyroscope configuration featuring a
macroscopic Sagnac area of $11$~cm$^2$.  We achieve a short-term
rotation stability of $100$~nrad.s$^{-1}/\sqrt{\rm Hz}$, and a
long-term stability as low as $1$~nrad.s$^{-1}$ after $10^4$ s of
integration, setting the record of all atom gyroscopes.

\section{Experimental Setup}
\begin{figure}
\centering
\includegraphics[width=8.6cm, page=2, trim = 5.5cm 5cm 3cm 4cm, clip]{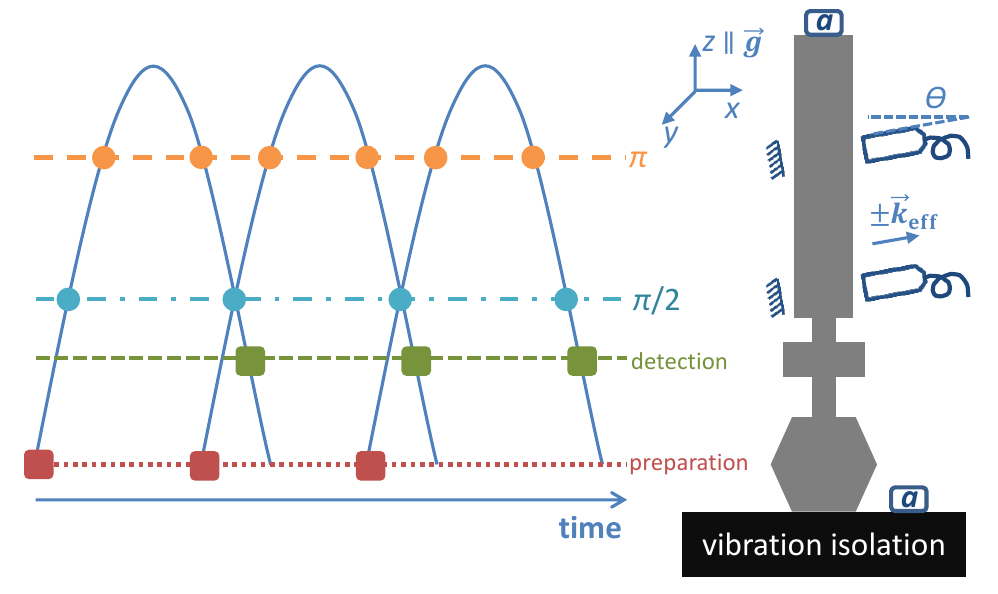}
\caption{\label{fig:fountain} Operation principle and setup of the
  continuous cold-atom gyroscope.  }
\end{figure}

We realize a light-pulse gyroscope in a cesium fountain, see
Fig.~\ref{fig:fountain}.  Counter-propagating Raman beams coupling the
$|F=4, m_F=0\rangle$ and $|F=3, m_F=0\rangle$ clock states are used to
split, deflect and recombine the free-falling cold atoms\footnote{We
  focus on the inertial measurement and analysis in this paper.
  Details of the atom preparation and detection can be found
  in~\cite{DuttaContinuous2016}.  }.  With four light pulses
($\pi/2$-$\pi$-$\pi$-$\pi/2$), the two arms of the interferometer
enclose a physical area up to $11$~cm$^2$, representing a $27$-fold
increase with respect to previous
experiments~\cite{berg_composite-light-pulse_2015}.  This gives rise
to a rotation phase shift $\Phi_\Omega$ according to the Sagnac
effect~\cite{Sagnac1913}, given by
\begin{equation}
  \Phi_\Omega = \frac{1}{2} \vec{k}_{\rm eff} \cdot \bigg( \vec{g}
  \times \vec{\Omega} \bigg) T^3,
\end{equation}
where $\vec{k}_{\rm eff}$ is the two-photon momentum transfer,
$\vec{g}$ is the gravitational acceleration, $\vec{\Omega}$ is the
rotation rate, and $T$ is half the interferometric time.  Following
the atom juggling methods initially introduced to measure collisional
shifts in fountain clocks~\cite{Legere1998}, we implement a sequence
of joint interrogation of successive atom clouds as described
in~\cite{Meunier2014}.  In other words, each $\pi/2$ Raman pulse is
common to the two adjacent interferometer sequences, setting the cycle
time $T_c$ equal to the total interrogation time $2T$.

The interrogation light contains two frequencies, each addressing one
of the two clock states.  The counter-propagating configuration is
achieved by means of retroreflecting the incoming beam (see
Fig.\ \ref{fig:fountain}), so that two configurations are possible,
transferring opposite momenta (denoted as $\pm \vec{k}_{\rm eff}$) to
the atoms.  The degeneracy of these two configurations is lifted by
tilting the Raman beams by an angle of inclination $\theta \simeq
4^\circ$, as the Doppler effect associated with the vertical velocity
of the atoms shifts the resonance frequency of the stimulated Raman transition in the opposite
directions.  The joint interrogation simultaneously addresses two atom
clouds with opposite vertical velocity, thus alternating between the
$\pm \vec{k}_{\rm eff}$ configurations.

Increasing the sensitivity of such AI-based inertial sensors
necessarily comes at the cost of an increased sensitivity to the
vibration noise, as a consequence of the Equivalence Principle.
Noninertial effects such as the Raman laser phase noise and light
shift also contribute to the inerferometer output.  We can thus
breakdown the interferometric phase into rotation phase, vibration
phase and noninertial phase, i.e.\ $\Delta \Phi = \Phi_\Omega + \delta
\Phi_{\rm vib} + \delta \Phi_0$.  Vibration noise has a strong impact
on our setup.  A vibration isolation platform reduces the effect of
the ground vibration $\gtrsim 1$~Hz to an rms AI phase noise of
about $2.5$~rad for $2T=800$~ms.  Since the vibration noise spans
several interferometer fringes, auxiliary inertial sensors are
necessary to recover the signal.

\begin{figure}
\centering
\includegraphics{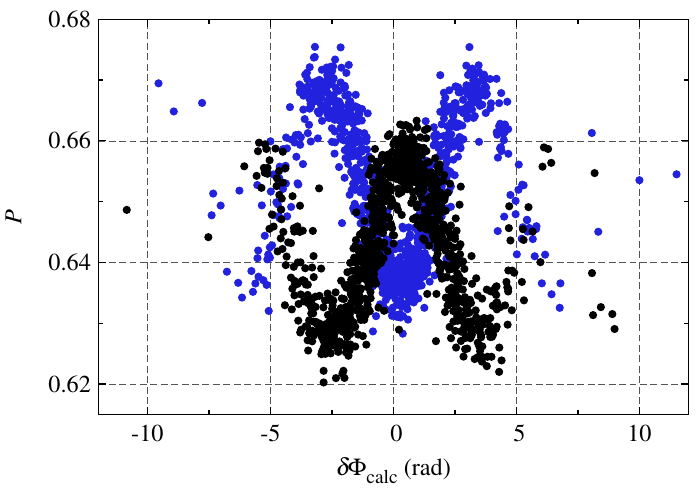}
\caption{\label{fig:fringe} Measured probability $P$ versus calculated
  vibration phase $\delta \Phi_{\rm calc}$ for the $\pm \vec{k}_{\rm
    eff}$ configurations.  }
\end{figure}

We use two commercial accelerometers (marked `{\em a}' in
Fig.\ \ref{fig:fountain}) to record and correct the vibration noise.
The acquired acceleration signal is weighted using the transfer
function~\cite{CheinetMeasurement2008} in order to compute the vibration phase.
Fig.~\ref{fig:fringe} shows the measured probability of transition $P$
versus the calculated vibration phase $\delta \Phi_{\rm calc}$ for the
$\pm \vec{k}_{\rm eff}$ configurations.  Despite the overwhelmingly
large vibration noise, the inertial stability of our gyroscope is
given by the horizontal scatter of the fringes.  This will be
evaluated in the following section.

\section{Stability of Rotation Measurement}
We divide a data set into packets of 40 points.  As the data
alternates between the the $\pm \vec{k}_{\rm eff}$ configurations, 20
points of each configuration are used to fit a sinusoidal model,
\begin{equation}
  P = P_0 + A \cos \big( \delta \Phi_{\rm calc} + \Phi^{(\pm)} \big),
\end{equation}
where $P_0$ is the offset of the interferometric signal, $A$ is the
fringe amplitude, and the phase offset is given by $\Phi^{(\pm)} = \pm
\Phi_\Omega + \delta \Phi_0$.  This yields a rotation phase
$\Phi_\Omega \equiv (\Phi^{(+)} - \Phi^{(-)})/2 \Mod{\pi}$.  All fitting parameters
$P_0$, $A$ and $\Phi^{(\pm)}$ were constrained loosely in order to
avoid cross talk between phase noise and probability or amplitude
noise.  The convergence of the fit routine is ensured by the large
span of the vibration phase.

\begin{figure}
\centering
\includegraphics{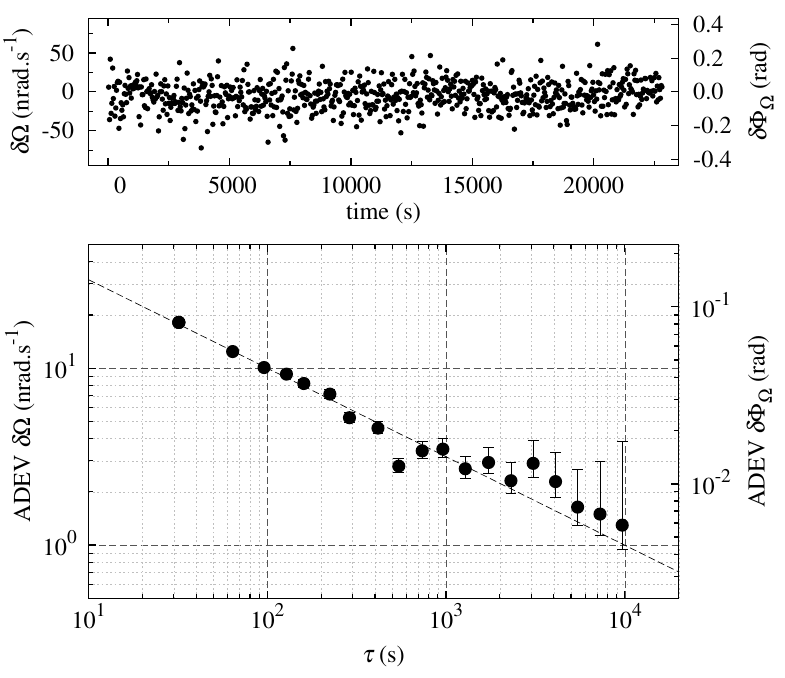}
  \caption{\label{fig:rotation} Top: The time sequence of rotation
    rate measurement around its mean value.  The equivalent
    fluctuation of rotation phase is shown on the right axis.  Bottom:
    ADEV of the measurement.  The error bars indicate the $68\%$
    confidence interval, and the dashed line follows a $\tau^{-1/2}$
    scaling.  }
\end{figure}

Figure~\ref{fig:rotation}~(top) shows an uninterrupted measurement over
about 6 hours.  The Allan standard deviation (ADEV) of the rotation
rate sensitivity is shown in Fig.~\ref{fig:rotation}~(bottom).  As the ADEV
follows the $\tau^{-1/2}$ scaling, where $\tau$ is the integration
time, we obtain a short-term rotation sensitivity of
$100$~nrad.s$^{-1}/\sqrt{\rm Hz}$.  This establishes the best
performance among all cold-atom gyroscopes to
date~\cite{berg_composite-light-pulse_2015}, and represents a 30-fold
improvement compared to previous four-pulse
gyroscopes~\cite{Canuel2006,stockton_absolute_2011}.  Comparing the
normal and the continuous mode, the performance of our gyroscope
improves by about a factor $1.4$.  This is consistent with the speedup
of the cycling frequency~\footnote{The dead time in normal mode $T_D
= 0.8$~s by coincidence, so that the cycling frequency doubles
  when we operate our gyroscope in the continuous mode.  }.

Such a sensitivity is currently limited by the detection noise (about
$400$~mrad$/\sqrt{\rm Hz}$ for $A\simeq 2\%$).  This also bounds the
efficiency of the vibration correction protocol to about a factor $5$
in the present case.  The technical difficulties associated with the
joint operation (primarily light shift and contrast reduction due to
scattered light by the MOT) are assessed in~\cite{Meunier2014},
together with strategies for improvement.

Nevertheless, the long-term stability of our rotation rate measurement
reaches $1$~nrad.s$^{-1}$ after $10^4$~s of integration time.  This
represents the state of the art of all atom
gyroscopes~\cite{Durfee2006} (see~\cite{barrett_sagnac_2014} for a
recent review), and a 20-fold improvement from previous cold-atom
gyroscopes~\cite{berg_composite-light-pulse_2015,gauguet_characterization_2009}.  Such a stability is a direct
consequence of the macroscopic Sagnac area and the folded four-pulse
geometry, giving a $T^3$ dependence of the scale factor.  With a long
interrogation time, fluctuations of the atom cloud trajectories, a
known limit in previous
experiments~\cite{berg_composite-light-pulse_2015,gauguet_characterization_2009}, are scaled down for its linear
dependence in $T$.  One-photon light shift, a source of slow drift in
stability due to the drift of the power ratio of the Raman lasers, is
removed by combining the measurement from the $\pm \vec{k}_{\rm eff}$
configurations.

\begin{figure}
\centering
\includegraphics[width=5.7cm, trim = 4.cm 5cm 4cm 4cm, clip]{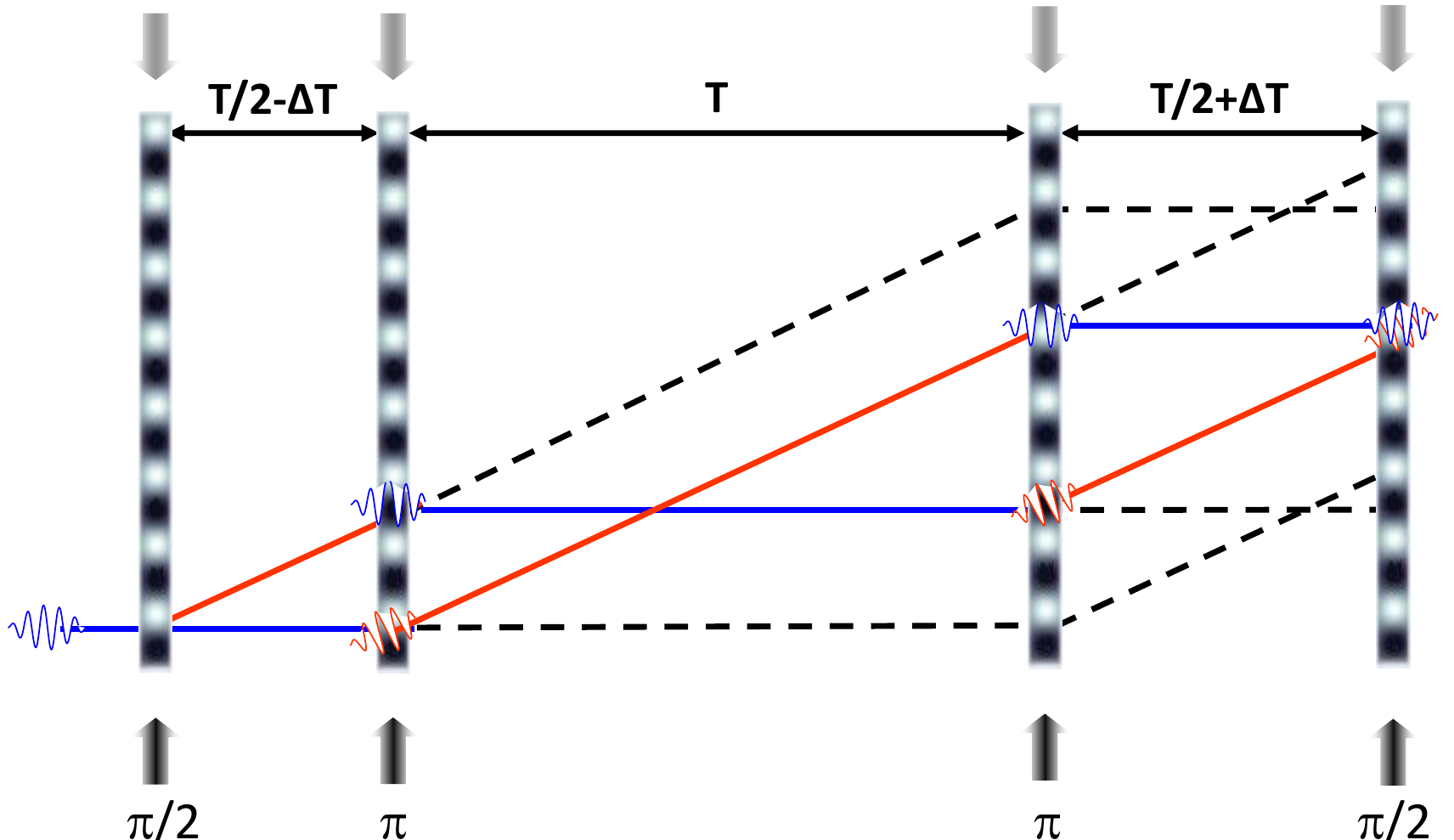}\\
\vspace{.2cm}\includegraphics{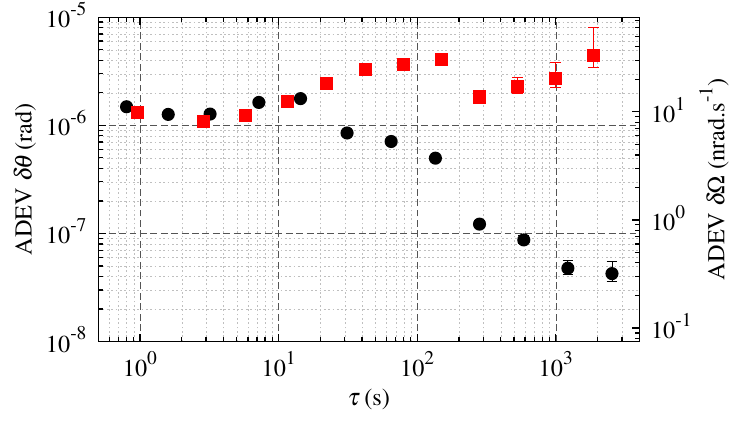}
  \caption{\label{fig:asym} Top: The space-time diagram of the
    asymmetric four-pulse interferometer.  Bottom: ADEV of the tilt
    measurement before (squares) and after (circles) active
    stablization.  }
\end{figure}

A symmetric four-pulse interferometer offers zero sensitivity to a DC
acceleration parallel to $\vec{k}_{\rm eff}$.  This however comes at
the expense of an enhanced probability noise in practice, as imperfect
$\pi$ pulses give rise to parasitic
interferometers~\cite{stockton_absolute_2011}.  We introduce a timing
asymmetry of $\Delta T = 300~\mu$s, see Fig.\ \ref{fig:asym}~(top), in
order to prevent the closure of the parasitic interferometers.  This
gives rise to a sensitivity to DC accelerations, $\Phi_{\rm DC} = 2
k_{\rm eff} T \Delta T g \sin \theta$.  In other words, a $1$~mrad
fluctuation of $\theta$ translates into a $\sim 70$~rad fluctuation of
$\Phi_{\rm DC}$.  We therefore stabilize the tilt of the experiment to
reduce these fluctuations.  A commercial tiltmeter is used to acquire
the tilt signal, the variations of which are compensated by a current
controlled magnetic actuator acting on the vibration platform.
Figure~\ref{fig:asym}~(bottom) shows the ADEV of $\delta \theta$ with and
without the tilt lock.  We stabilize $\delta \theta$ down to $\sim
4\times 10^{-8}$~rad, corresponding to a long-term stabilization of
$\Phi_{\rm DC}$ below $0.3$~nrad.s$^{-1}$ level after $2000$~s of
integration.  Alternating measurements between $\Delta T = \pm
300~\mu$s allowed us to verify that $\Phi_{\rm DC}$ does not impact
the stability of the rotation rate measurement.  We also monitor the
cross-axis tilt and observe a negligible phase drift due to the change
of the projection of the rotation vector on the interferometer area.

\section{Towards Quantum Noise Limited Atom Interferometers}
The continuous operation introduces phase correlations between
successive measurements.  This in principle allows faster noise
averaging following a $\tau^{-1}$ scaling in ADEV.  It has been
demonstrated on our setup in the clock mode~\cite{Meunier2014} where
the Dick effect from a degraded local oscillator is quickly reduced
with integration.

\begin{figure}
  \centering
  \includegraphics{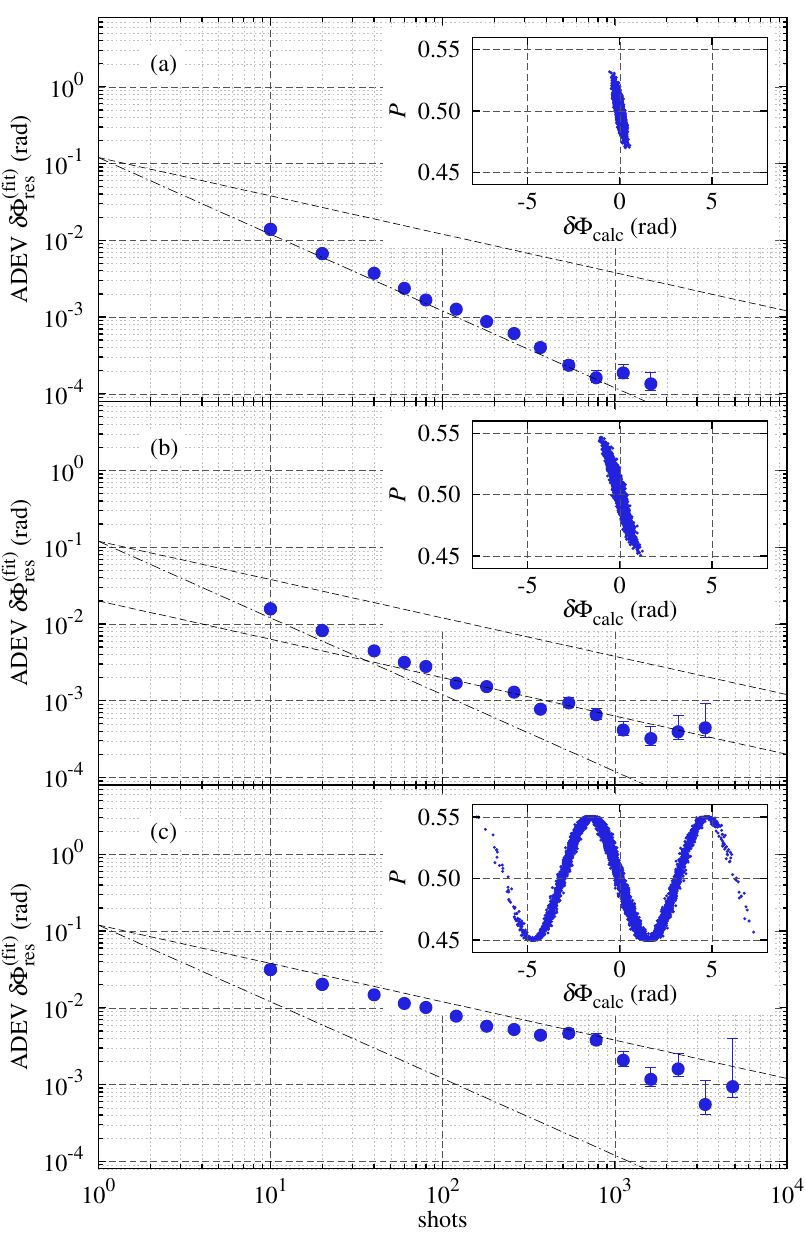}
  \caption{\label{fig:simu} Simulation of the gyroscope phase
    stability with increasing vibration noise.  (a) - (c) correspond
    to $\sigma_{\rm vib} = 0.13$~rad, $0.32$~rad, and $2.1$~rad added
    to a constant level of residual correlated noise $\sigma_{\rm res}
    = 120$~mrad.  The fast noise averaging following $\tau^{-1}$
    scaling (dash-dotted lines) is gradually lost for AIs that operate
    away from mid-fringe, recovering the $\tau^{-1/2}$ scaling (dashed
    lines).  The insets show the simulated $P$ versus $\delta
    \Phi_{\rm calc}$.  }
\end{figure}

In order to demonstrate the same $\tau^{-1}$ scaling in our inertial
measurements, we need to reduce the uncorrelated detection noise, and
to operate our AI at mid-fringe in order to preserve the maximal
sensitivity, i.e.\ $|dP/d\Delta \Phi| = A$.  This is confirmed by a
simulation of the ADEV for different levels of vibration noise, which
is corrected by auxiliary sensors.  The residual phase noise $\delta
\Phi_{\rm res}$ (including the inertial noise not corrected by the auxiliary sensors and some noninertial noise) is correlated
between successive shots.  Its rms $\sigma_{\rm res} = 120$~mrad is
kept constant in all three cases.  The vibration noise calculated from the auxiliary sensor signals is generated randomly, with an rms of $\sigma_{\rm calc} = 0.13$~rad, $0.32$~rad, and $2.1$~rad.
For $P_0 = 0.5$ and $A = 5\%$, we compute $P = P_0 - A \sin
\big(\delta \Phi_{\rm calc} + \delta \Phi_{\rm res}\big)$ to simulate
the AI operation.  Fitting $P$ versus $\delta \Phi_{\rm calc}$ using
10-point packets yields $\delta \Phi_{\rm res}^{\rm (fit)}$, similar
to our data analysis procedure.  The ADEV of $\delta \Phi_{\rm
  res}^{\rm (fit)}$ is shown in Fig.~\ref{fig:simu}, indicating a loss
of the $\tau^{-1}$ scaling when the vibration noise brings the AI out
of the linear regime.  Note that fringe fitting is equivalent to
linear regression as long as the AI remains at mid-fringe, see
Fig.\ \ref{fig:simu}~(a).

We can retain a mid-fringe operation using a real-time compensation of
the vibration noise, first demonstrated on an atom
gravimeter~\cite{lautier_hybridizing_2014}.  A phase jump of the
interrogation laser right before the end of the interferometer
sequence can cancel the vibration phase and reduce the span of the
interferometric phase.  Alternatively, we can implement a more elaborated protocol using quantum weak measurement, as shown in the clock mode in~\cite{KohlhaasPhase2015}.  Assuming a quantum projection noise limited
detection with $10^6$ atoms and $A=10\%$, a rotation sensitivity below
$1\times 10^{-10}$~rad.s$^{-1}$ in a few 100~s is accessible with our
setup.

\section{Conclusion}
The continuous operation of our atom gyroscope allows us to improve
the stability of the rotation rate measurement without loss of
inertial information.  We report $100$~nrad.s$^{-1}/\sqrt{\rm Hz}$
rotation sensitivity, and a stability of $1$~nrad.s$^{-1}$ after
$10^4$~s of integration.  This is well within the specifications of a
strategic grade gyroscope ($<4$~nrad.s$^{-1}$
stability~\cite{lefevre_fiber-optic_2014}), making AI more attractive for
inertial navigation.  We also foresee applications in geodesy and
geophysics, where seismic signals in the a few mHz to 10s of Hz
frequency band could be accessible with such inertial sensors.  AI
operating in continuous mode are also useful in the search of
time-dependent signals such as gravitational
waves~\cite{Dimopoulos2008PRD,Chaibi2016}.

\section*{Acknowledgement}
We acknowledge the financial support from D\'el\'egation G\'en\'erale de l'Armement (contract No. 2010.34.0005), Centre National d'Etudes Saptiales, Institut Francilien de Recherche sur les Atomes Froids, the Action Sp\'ecifique du CNRS Gravitation, R\'ef\'erences, Astronomie et M\'etrologie and Ville de Paris (project HSENS-MWGRAV). I.D. was supported by CNES and FIRST-TF (ANR-10-LABX-48-01), D.S. by DGA and B.F. by FIRST-TF.  

\bibliographystyle{IEEEtran}
\bibliography{IEEEabrv,EFTFproceeding}

\begin{thebibliography}{10}
\providecommand{\url}[1]{#1}
\csname url@samestyle\endcsname
\providecommand{\newblock}{\relax}
\providecommand{\bibinfo}[2]{#2}
\providecommand{\BIBentrySTDinterwordspacing}{\spaceskip=0pt\relax}
\providecommand{\BIBentryALTinterwordstretchfactor}{4}
\providecommand{\BIBentryALTinterwordspacing}{\spaceskip=\fontdimen2\font plus
\BIBentryALTinterwordstretchfactor\fontdimen3\font minus
  \fontdimen4\font\relax}
\providecommand{\BIBforeignlanguage}[2]{{%
\expandafter\ifx\csname l@#1\endcsname\relax
\typeout{** WARNING: IEEEtran.bst: No hyphenation pattern has been}%
\typeout{** loaded for the language `#1'. Using the pattern for}%
\typeout{** the default language instead.}%
\else
\language=\csname l@#1\endcsname
\fi
#2}}
\providecommand{\BIBdecl}{\relax}
\BIBdecl

\bibitem{Canuel2006}
B.~Canuel, F.~Leduc, D.~Holleville, A.~Gauguet, J.~Fils, A.~Virdis, A.~Clairon,
  N.~Dimarcq, C.~J. Bord\'e, A.~Landragin, and P.~Bouyer, ``Six-axis inertial
  sensor using cold-atom interferometry,'' \emph{Phys. Rev. Lett.}, vol.~97, p.
  010402, Jul 2006.

\bibitem{Geiger2011}
R.~Geiger, V.~Menoret, G.~Stern, N.~Zahzam, P.~Cheinet, B.~Battelier,
  A.~Villing, F.~Moron, M.~Lours, Y.~Bidel, A.~Bresson, A.~Landragin, and
  P.~Bouyer, ``Detecting inertial effects with airborne matter-wave
  interferometry,'' \emph{Nat Commun}, vol.~2, p. 474, Sep. 2011.

\bibitem{stockton_absolute_2011}
J.~K. Stockton, K.~Takase, and M.~A. Kasevich, ``Absolute {Geodetic} {Rotation}
  {Measurement} {Using} {Atom} {Interferometry},'' \emph{Phys. Rev. Lett.},
  vol. 107, no.~13, p. 133001, Sep. 2011.

\bibitem{Gillot2014}
P.~Gillot, O.~Francis, A.~Landragin, F.~P.~D. Santos, and S.~Merlet,
  ``Stability comparison of two absolute gravimeters: optical versus atomic
  interferometers,'' \emph{Metrologia}, vol.~51, no.~5, p. L15, 2014.

\bibitem{Geiger2015}
{R. Geiger et al}, ``Matter-wave laser interferometric gravitation antenna
  (miga): New perspectives for fundamental physics and geosciences,'' in
  \emph{Proceedings of the 50th Rencontres de Moriond}, 2015.

\bibitem{Freier2015}
C.~Freier, M.~Hauth, V.~Schkolnik, B.~Leykauf, M.~Schilling, H.~Wziontek, H.-G.
  Scherneck, J.~M\"uller, and A.~Peters, ``{Mobile quantum gravity sensor with
  unprecedented stability},'' in \emph{Proceedings for the 8th Symposium on
  Frequency Standards and Metrology}, 2015.

\bibitem{RosiMeasurement2015}
G.~Rosi, L.~Cacciapuoti, F.~Sorrentino, M.~Menchetti, M.~Prevedelli, and G.~M.
  Tino, ``Measurement of the gravity-field curvature by atom interferometry,''
  \emph{Phys. Rev. Lett.}, vol. 114, p. 013001, Jan 2015.

\bibitem{AltschulQuantum2015}
B.~Altschul, Q.~G. Bailey, L.~Blanchet, K.~Bongs, P.~Bouyer, L.~Cacciapuoti,
  S.~Capozziello, N.~Gaaloul, D.~Giulini, J.~Hartwig, L.~Iessm, P.~Jetzer,
  A.~Landragin, E.~Rasel, S.~Reynaud, S.~Schiller, C.~Schubert, F.~Sorrentino,
  U.~Sterr, J.~D. Tasson, G.~M. Tino, P.~Tuckey, and P.~Wolf, ``Quantum tests
  of the einstein equivalence principle with the ste?quest space mission,''
  \emph{Advances in Space Research}, vol.~50, p. 501, 2015.

\bibitem{Dimopoulos2008PRD}
S.~Dimopoulos, P.~W. Graham, J.~M. Hogan, M.~A. Kasevich, and S.~Rajendran,
  ``Atomic gravitational wave interferometric sensor,'' \emph{Phys. Rev. D},
  vol.~78, p. 122002, Dec 2008.

\bibitem{Chaibi2016}
W.~Chaibi, R.~Geiger, B.~Canuel, A.~Bertoldi, A.~Landragin, and P.~Bouyer,
  ``Low frequency gravitational wave detection with ground-based atom
  interferometer arrays,'' \emph{Phys. Rev. D}, vol.~93, p. 021101, Jan 2016.

\bibitem{Clade2009}
P.~Clad\'e, S.~Guellati-Kh\'elifa, F.~Nez, and F.~Biraben, ``Large momentum
  beam splitter using bloch oscillations,'' \emph{Phys. Rev. Lett.}, vol. 102,
  p. 240402, Jun 2009.

\bibitem{Chiow2011}
S.-w. Chiow, T.~Kovachy, H.-C. Chien, and M.~A. Kasevich,
  ``$102\ensuremath{\hbar}k$ large area atom interferometers,'' \emph{Phys.
  Rev. Lett.}, vol. 107, p. 130403, Sep 2011.

\bibitem{Debs2011}
J.~E. Debs, P.~A. Altin, T.~H. Barter, D.~D\"oring, G.~R. Dennis, G.~McDonald,
  R.~P. Anderson, J.~D. Close, and N.~P. Robins, ``Cold-atom gravimetry with a
  bose-einstein condensate,'' \emph{Phys. Rev. A}, vol.~84, p. 033610, Sep
  2011.

\bibitem{Leroux2010}
I.~D. Leroux, M.~H. Schleier-Smith, and V.~Vuleti\ifmmode~\acute{c}\else
  \'{c}\fi{}, ``Implementation of cavity squeezing of a collective atomic
  spin,'' \emph{Phys. Rev. Lett.}, vol. 104, p. 073602, Feb 2010.

\bibitem{Hosten2016}
O.~Hosten, N.~J. Engelsen, R.~Krishnakumar, and M.~A. Kasevich, ``Measurement
  noise 100 times lower than the quantum-projection limit using entangled
  atoms,'' \emph{Nature}, vol. 529, no. 7587, pp. 505--508, Jan. 2016.

\bibitem{Fang2016}
B.~Fang, I.~Dutta, P.~Gillot, D.~Savoie, J.~Lautier, B.~Cheng, C.~L.
  Garrido~Alzar, R.~Geiger, S.~Merlet, F.~Pereira Dos~Santos, and A.~Landragin,
  ``Metrology with atom interferometry: Inertial sensors from laboratory to
  field applications,'' in \emph{Proceedings for the 8th Symposium on Frequency
  Standards and Metrology}, 2016.

\bibitem{jekeli_navigation_2005}
C.~Jekeli, ``Navigation {Error} {Analysis} of {Atom} {Interferometer}
  {Inertial} {Sensor},'' \emph{Navigation}, vol.~52, no.~1, pp. 1--14, Mar.
  2005.

\bibitem{Schreiber2013}
K.~U. Schreiber and J.-P.~R. Wells, ``Invited review article: Large ring lasers
  for rotation sensing,'' \emph{Rev. Sci. Instrum.}, vol.~84, no.~4, 2013.

\bibitem{SantarelliFrequency1998}
G.~Santarelli, A.~Audoin, C.and~Makdissi, P.~Laurent, G.~J. Dick, and
  A.~Clairon, ``Frequency stability degradation of an oscillator slaved to a
  periodically interrogated atomic resonator,'' \emph{IEEE Trans. Ultrason.
  Ferroelectr. Freq. Control}, vol.~45, pp. 887--894, 1998.

\bibitem{DuttaContinuous2016}
I.~Dutta, D.~Savoie, B.~Fang, B.~Venon, C.~L. Garrido~Alzar, R.~Geiger, and
  A.~Landragin, ``Continuous cold-atom inertial sensor with $1$~nrad.s$^{-1}$
  rotation stability,'' \emph{Phys. Rev. Lett.}, vol. 116, p. 183003, 2016.

\bibitem{berg_composite-light-pulse_2015}
P.~Berg, S.~Abend, G.~Tackmann, C.~Schubert, E.~Giese, W.~P. Schleich, F.~A.
  Narducci, W.~Ertmer, and E.~M. Rasel, ``Composite-{Light}-{Pulse} {Technique}
  for {High}-{Precision} {Atom} {Interferometry},'' \emph{Phys. Rev. Lett.},
  vol. 114, no.~6, p. 063002, Feb. 2015.

\bibitem{Sagnac1913}
G.~Sagnac, ``{L'\'{e}ther lumineux d\'{e}montr\'{e} par l'effet du vent relatif
  d'\'{e}ther dans un interf\'{e}rom\`{e}tre en rotation uniforme},'' \emph{C.
  R. Acad. Sci. (Paris)}, vol. 157, p. 708, 1913.

\bibitem{Legere1998}
R.~Legere and K.~Gibble, ``Quantum scattering in a juggling atomic fountain,''
  \emph{Phys. Rev. Lett.}, vol.~81, pp. 5780--5783, Dec 1998.

\bibitem{Meunier2014}
M.~Meunier, I.~Dutta, R.~Geiger, C.~Guerlin, C.~L. Garrido~Alzar, and
  A.~Landragin, ``Stability enhancement by joint phase measurements in a single
  cold atomic fountain,'' \emph{Phys. Rev. A}, vol.~90, p. 063633, Dec 2014.

\bibitem{CheinetMeasurement2008}
P.~Cheinet, B.~Canuel, F.~Pereira Dos~Santos, A.~Gauguet, F.~Leduc, and
  A.~Landragin, ``Measurement of the sensitivity function in time-domain atomic
  interferometer,'' \emph{IEEE Trans. on Instrum. Meas.}, vol.~57, p. 1141,
  2008.

\bibitem{Durfee2006}
D.~S. Durfee, Y.~K. Shaham, and M.~A. Kasevich, ``Long-term stability of an
  area-reversible atom-interferometer sagnac gyroscope,'' \emph{Phys. Rev.
  Lett.}, vol.~97, p. 240801, Dec 2006.

\bibitem{barrett_sagnac_2014}
B.~Barrett, R.~Geiger, I.~Dutta, M.~Meunier, B.~Canuel, A.~Gauguet, P.~Bouyer,
  and A.~Landragin, ``The {Sagnac} effect: 20 years of development in
  matter-wave interferometry,'' \emph{Comptes Rendus Physique}, vol.~15,
  no.~10, pp. 875--883, Dec. 2014.

\bibitem{gauguet_characterization_2009}
A.~Gauguet, B.~Canuel, T.~Lévèque, W.~Chaibi, and A.~Landragin,
  ``Characterization and limits of a cold-atom {Sagnac} interferometer,''
  \emph{Phys. Rev. A}, vol.~80, no.~6, p. 063604, Dec. 2009.

\bibitem{lautier_hybridizing_2014}
J.~Lautier, L.~Volodimer, T.~Hardin, S.~Merlet, M.~Lours, F.~{Pereira Dos
  Santos}, and A.~Landragin, ``Hybridizing matter-wave and classical
  accelerometers,'' \emph{Appl. Phys. Lett.}, vol. 105, no.~14, p. 144102, Oct.
  2014.

\bibitem{KohlhaasPhase2015}
R.~Kohlhaas, A.~Bertoldi, E.~Cantin, A.~Aspect, A.~Landragin, and P.~Bouyer,
  ``Phase locking a clock oscillator to a coherent atomic ensemble,''
  \emph{Phys. Rev. X}, vol.~5, p. 021011, Apr 2015.

\bibitem{lefevre_fiber-optic_2014}
H.~C. Lef\'evre, ``The fiber-optic gyroscope, a century after {Sagnac}'s
  experiment: {The} ultimate rotation-sensing technology?'' \emph{Comptes
  Rendus Physique}, vol.~15, no.~10, pp. 851--858, Dec. 2014.

\end{thebibliography}

\end{document}